\documentclass[epj]{svjour}

\usepackage{bm}% bold math

\usepackage{amsfonts}

\usepackage{amscd}
\usepackage{color}
\usepackage{graphicx}

\usepackage{multirow}

\usepackage[colorlinks = true, linkcolor = blue, citecolor = blue]{hyperref}

\usepackage{amsmath}
\usepackage{appendix}

\begin{document}

\titlerunning{Quantum high-frequency conductivity oscillations...}

\title{Quantum high-frequency conductivity oscillations
in graphene multilayers and nodal semimetals in a tilted magnetic
field}

%\title{Magneto-angular oscillations of high-frequency conductivity
%in graphene multilayers and nodal semimetals}

%\title{Quantum cyclotron resonance of Dirac electrons in graphene multilayers and nodal semimetals in a tilted magnetic field}

%\title{Angular dependence of the cyclotron resonance peak amplitude
%of Dirac electrons in graphene multilayers and nodal semimetals}

\author{Juan C. Medina Pantoja\inst{1}  \and  Juan Sotelo-Campos\inst{1} \and Igor V. Kozlov\inst{2}  }
\institute{Laboratorio de Ciencias de los Materiales, Facultad de
Ciencias y Filosofia, Universidad Peruana Cayetano Heredia, \\
Av. Honorio Delgado 430, 31 Lima, Peru \and B. Verkin Institute
for Low Temperature Physics and Engineering of the National
Academy of Sciences of Ukraine, \\
47 Nauky Ave., 61103 Kharkiv, Ukraine \\
\email{kozlov@ilt.kharkov.ua} }

\date{}

\abstract{A new type of angular oscillations of the high-frequency
conductivity for conductors with a band-contact line has been
predicted. The effect is caused by groups of charge carriers near
the self-intersection points of the Fermi surface, where the
electron energy spectrum is near-linear and can be described by
anisotropic Dirac cone model. The amplitude of the resonance peaks
satisfies the simple sum rule. The ease in changing the degree of
anisotropy of the Dirac cone due to the angle of inclination of
the magnetic field makes the considered type of oscillations
attractive for experimental observation of relativistic  effects.
\\
 }


\maketitle

\section{Introduction}
\label{sec:level1}
Recently, there has been growing interest in the study of nodal
semimetals having band-contact lines. First principles
colculations indicates the existance of ring-shaped nodal lines in
Ca, Sr, Yt \cite{Hirayama2016}. Also, the topological transition 3
$ \frac{1}{2} $ kind is known for the conductors with band-contact
line and thus possible in graphite conductors family, Be, Mg, Zn,
Cd, Al and other materials \cite{Mikitik2014}. Usual graphite have
nodal lines \cite{Mikitik}.

In this paper we call attention to the effects of anisotropic
Dirac cones without an inversion center (tilted Dirac cones) in
nodal semimetals. The Hamiltonian, corresponding to the linear
energy spectrum of Dirac-type charge carriers has the form
\cite{Sari,Morinari,Sari2}:
\begin{equation}
\epsilon(p_x , p_y) = v_0 ( \alpha \sigma_x p_x + \sigma_y p_y +
\eta p_y ) \label{spektr}
\end{equation}
where the absence of an inversion centers $ \eta \neq 0 $ is
either a consequence of the internal symmetry of the conductor or
may be achieved artificially, e.g. in strained graphene or in a
problem of Dirac electron drift in crossed electric and magnetic
fields \cite{Sari}. The so-called "tilt" $ \eta $ can describe the
relativistic effects \cite{goerbig-obzor}. Furthermore, the
"collapse" ($ |\eta| > 1 $) of the Hamiltonian (\ref{spektr}) is
naturally explained in terms of relativistic rotations ("Lorentz
boosts") \cite{Collaps}.

The implementation of the Hamiltonian (\ref{spektr}) for graphene
requires a relatively strong electric fields $ \sim 10^6 V/m $ and
relatively large strain values in the sample $ \sim 10 \% $
\cite{goerbig-obzor}. In case of natural anisotropy, particulary
in the compound $ \alpha-(BEDT-TTF)_2 I_3 $
\cite{Sari,Morinari,Morinari2,Morinari3,Morinari4}, changing of
parameters of the electron energy spectrum is difficult, since the
latter is due to the intrinsic properties of the conductor.
Thus the experimental observation of effects that require a
parameter $ \eta $ to be changed is associated with certain
difficulties in these conductors. All  the above mentioned
conductors have a pronounced two-dimensional nature.

At the same time physical phenomena characteristic of the
Hamiltonian (\ref{spektr}), will take place in nodal semimetals
%the family of graphite conductors
near the self-intersection points of Fermi surfaces. It can be
noticed that in a tilted magnetic field, the electron energy
spectrum in Larmor orbit's plane will be given by the model
(\ref{spektr}), where the value of $ \eta $ which determines the
degree of anisotropy of the electron energy spectrum, can be
easily changed by simply changing the tilt angle of a quantizing
magnetic field.
The attractiveness of the graphite and its derivatives is
 determined by the fact that for simple
chemical compounds the high purity of the conductor required for
the observation of high harmonics of the quantum cyclotron
resonance \cite{Orlita} can be achieved easier.
%

%The model (\ref{spektr}) is applicable for describing the electron
%energy spectrum in Larmor orbit's plane for various conductors
%with a band-contact line.
%
%Recently, there has been growing interest in nodal semimetals
%having ring-shaped band-contact lines. A number of compounds have
%a similar feature of the energy spectrum. { \color{red} Indeed,
%first principles colculations indicates the existance of
%ring-shaped nodal lines in Ca, Sr, Yt \cite{Hirayama2016}. Also,
%the topological transition 3 $ \frac{1}{2} $ kind is known for the
%conductors with band-contact line and thus possible in graphite
%conductors family, Be, Mg, Zn, Cd, Al and other materials
%\cite{Mikitik2014}. }

%In this work we have theoretically observed the resonant effects for

The goal of the present work is new oscillation phenomena caused
by tilted Dirac cone effects in conductors having band-contact
(nodal) lines. In Section "Model" we choose the model of the
electron energy spectrum. The model has the qualitative accordance
with the Fermi surface of a number of nodal semimetals. The
conditions limiting the applicability of the model are given. In
Section "Conductivity tensor" a new type of angular oscillations
of the high-frequency conductivity for conductors with nodal lines
has been predicted. The physical mechanism of these oscillations
is explained. Section "Sum rule" shows, that the amplitude of the
resonance peaks satisfies the simple sum rule or the "magic square
rule", which follows directly from properties of Pauli matrices.
In Section "Conclusions" the paper is summarized and concluded.
The possibility to observe the predicted oscillation effects is
discussed. We provide a brief overview of articles related to the
results of the present work.

\section{Model}
The model of graphene multilayers with a crystal lattice of AA
type stacking is convenient for the observation of the resonant
effects near the Dirac cone, since the character of the energy
spectrum of the charge carriers can be considered to be linear in
a broad range of energies \cite{Abrikosov,lobato}. AA stacked
graphite is unstable and cannot exist under natural conditions.
Although the nanoparticles with the number of layers of about ten
grown on the border of the diamond may be available for direct
observation \cite{diamond}.  Nevertheless, the energy spectrum of
AA type graphite is widely used in theoretical works as the
simplest and the most convenient theoretical model due to its
characteristics, such as layering and Dirac energy spectrum of
charge carriers near the Fermi surface (see the work)
\cite{Abrikosov} (see also Ref.\cite{KozlovMedina}).
The Hamiltonian of low-energy charge carriers carriers
corresponding to the model has the form:
\begin{equation}
H ({\bf p}) = v_\| ( \sigma_x p_x + \sigma_y p_y ) - 2t \cos
(\frac{a_z p_z}{\hbar}) , \label{spektr-abrikosov}
\end{equation}
 where $ a_z $ is the interlayer distance and t is the overlap integral of the wave
functions in adjacent layers, that we consider to be positive.
This model was proposed for the conductors with a graphitelike
energy spectrum (\ref{spektr-abrikosov}) in Ref.\cite{Abrikosov},
where a linear magnetoresistance of  a  layered conductor with a
small overlap integral t was investigated.

One can also easily see the qualitative accordance of the model
(\ref{spektr-abrikosov}) with a fragment of the Fermi surface of a
number of nodal semimetals near the point of self-intersection of
Fermi surfaces (see. Fig.1 of Ref. \cite{Mikitik2014}). In
particular the topological transition of 3 $ \frac{1}{2} $ kind
\cite{Mikitik2014} occurs when $ \epsilon_F = \pm 2t $ for the
model (\ref{spektr-abrikosov}).

This model of the electron energy spectrum can be also suitable
for a number of graphite intercalates with AA type stacking of
graphene layers \cite{Dresselhaus}. For example, recent ARPES
studies have reported about the direct observation of a linear
energy spectrum of charge carriers in $ KC_8 $ compounds. Along
with the observed data concerning the traditional quantum
oscillation effects, the ARPES results reveal the applicability of
the Dirac cone model for the energy spectrum of the charge
carriers in conductors of this type \cite{ARPES}. The dependence
of the energy of the charge carriers on the momentum components in
the plane of the layers with a good degree of accuracy can be
considered to be linear in the energy area of the order of
fractions of ev, which is significantly higher than in graphite
(several mevs) \cite{wang,ARPES}. The Fermi velocity in the layers
plane $ v_F \approx (0.82 - 0.97) \times 10^6 m/s $ (see for
example \cite{Orlita}), i.e. is close to the value of the Fermi
velocity of conduction electrons in graphene. Unfortunately, a
strong shift of the Fermi level is often observed in intercalated
graphite. Therefore the Dirac singularity can be deep below the
Fermi level, that takes place for intercalation by alkali metals
in particular. Nevertheless, the wide variety of intercalated
graphite compounds gives the possibility to observe the effects of
an anisotropic Dirac cone for the other members of this family of
compounds. The model (\ref{spektr-abrikosov}) was later used in
Ref.\cite{KozlovMedina} to study the quantum cyclotron resonance
in the case of not so high frequencies $ \hbar\omega < \epsilon_1
$, where $ \epsilon_1 $ is the energy difference between the
zeroth and first Landau levels, when the influence of the
electron-hole transitions can be neglected.

In a tilted magnetic field $ {\bf B}=(0,B_0 \sin \theta,B_0 \cos
\theta) $, near the self-intersection point of the Fermi surface $
{\bf p } = (0,0,p_{z0}) $, the dependence of the charge carriers
energy on the components of the momentum in Larmor orbit's plane
can be described by the expression (\ref{spektr}) with the
parameter values
\begin{eqnarray}
\eta = - \frac{v_\bot}{v_\|} \tan \theta, \qquad \qquad v_0 = v_\|
\cos \theta,
\nonumber \\
\alpha = \frac{1}{\cos \theta}, \qquad v_\bot =
\frac{2ta_z}{\hbar} \sin \frac{a_z p_{z0}}{\hbar},
\label{const-theta-2}
\end{eqnarray}
( $ v_\bot $ is the Fermi velocity of conduction electrons along
the normal to the layers. ) We assume that the inequality $ |
\epsilon_F | < 2t $ holds, in which the Fermi surface has
self-intersection points. We concentrate on the frequency region $
\hbar\omega
> \epsilon_1 $, so that the representation of the cyclotron
resonance is determined by electron-hole transitions. The quantum
cyclotron resonance and the classical contribution to the high
frequency conductivity in the frequency region $ \hbar\omega <<
\epsilon_1 $, where the infuence of electron-hole transition is
negligible, have already been considered in
Ref.\cite{KozlovMedina} for the case of the magnetic field normal
to the layers. The deviation from the linear dependence
(\ref{spektr}) can be neglected for Landau levels with $\epsilon_n
\sim \hbar\omega$ , (\ref{En}) for angles $ \theta $ of the
magnetic field $ \mathbf{B} $ satisfying the following inequality,
which is considered to hold from now onwards:
\begin{equation}
\epsilon_1 < \hbar\omega << min\{\epsilon_1 / \eta , \ (2t \pm
\epsilon_F)\} \label{omega-usl} .
\end{equation}
We only consider the case of a sufficiently large relaxation time
$ \tau $ and relatively low temperatures $ T $:
\begin{equation}
\frac{2t \pm \epsilon_F}{\hbar\omega} << \omega\tau << \frac{\hbar
v_0}{a_z T \tan \; \theta} \label{usl-T} .
\end{equation}
The right side of the inequality allows us to neglect the
deviation from the linear dependence (\ref{spektr}) in the region
of temperature smearing of the Fermi surface
(\ref{spektr-abrikosov}), near the latter's points of
self-intersection.
Also later we will consider only the diagonal matrix elements of
the conductivity tensor $ \sigma_{ii} $ in the plane $
(\tilde{x},\tilde{y}) $ orthogonal to the vector {\bf B},
\begin{equation}
\tilde{x} = x, \; \tilde{y} = y \cos \theta - z \sin \theta, \;
\tilde{z} = z \cos \theta + y \sin \theta . \label{turn}
\end{equation}
Here in after the sign "tilde" will be used to denote the
components in the rotated coordinate system (\ref{turn}). While
calculating the conductivity tensor components, we will use
quantum kinetic equation in the relaxation time approximation.

\section{Conductivity tensor}
Within the relaxation time approximation, the conductivity tensor
$ \sigma_{ij} (\omega) $ can be written in the form:
\begin{eqnarray}
\sigma_{ij} (\omega) = \frac{2e^3 B}{(2\pi\hbar)^2 c}
\sum_{n,m=-\infty}^\infty \int d\tilde{p}_z \left( -\frac{f^0_n
(\tilde{p}_z) - f^0_m (\tilde{p}_z)}{E_n (\tilde{p}_z) - E_m
(\tilde{p}_z)} \right)
\nonumber  \\   \times
\frac{v^j_{mn} (\tilde{p}_z) v^i_{nm} (\tilde{p}_z)}{-i\omega +
\frac{i}{\hbar}(E_n (\tilde{p}_z) - E_m (\tilde{p}_z)) +
\frac{1}{\tau}}, \qquad \label{kin-ur}
\end{eqnarray}
where $ f_n^0 (\tilde{p}_z) = f^0 (E_n (\tilde{p}_z)) $ is the
Fermi-Dirac function, $ \tau $ is the relaxation time, $
v^{i,j}_{nm} $ are the matrix elements of the velocity operator
and $ \tilde{p}_z $ is the projection of the momentum vector onto
the magnetic field vector B.
The factor of 2 in the numerator is obtained from summation over
the conventional spin.
If the inequalities (\ref{omega-usl},\ref{usl-T}) are satisfied,
the energy levels for model (\ref{spektr-abrikosov}) can be
represented as
\begin{equation}
E_n (\tilde{p}_z) =  \epsilon_1 (\tilde{p}_z) sign(n) \sqrt{|n|} -
2t \cos \left(\frac{a_z \tilde{p}_z}{\hbar \, \cos \, \theta
}\right) + \delta E_n (\tilde{p}_z) . \label{En-full}
\end{equation}
Caused by the deviation from the model (\ref{spektr}) the
amendment $ |\delta E_n (\tilde{p}_z)| << \frac{1}{\tau} $, does
not significantly affect the position of the resonance peaks and
can be omitted. The energy of the first Landau level
\begin{equation}
\epsilon_1 (\tilde{p}_z) = v_\| \;  \sqrt{2 \cos \theta
\frac{eB\hbar}{ c} \lambda^3 (\tilde{p}_z)},
\end{equation}
where
\begin{equation}
\lambda (\tilde{p}_z) = \sqrt{ 1 - \left( \frac{2 t a_z}{\hbar
v_\|} \sin \left(\frac{a_z \tilde{p}_z}{\hbar \cos \theta}\right)
\tan \theta \right)^2 } ,
\end{equation}
can be easily obtained from the expression (\ref{En}) with the
energy spectrum parameters in (\ref{const-theta-2}) evaluated at
the point $ p_x=p_y=0, \; p_z=\tilde{p}_z / \cos \theta $, and is
given here for succession.

In the frequency region delimited by the inequalities
(\ref{omega-usl}, \ref{usl-T}), the conductivity oscillations will
be determined by the charge carriers near the self-intersection
points of the Fermi surface, for which the dependence of the
cyclotron frequency $ \hbar\Omega_n (\tilde{p}_z) = E_{n+1}
(\tilde{p}_z) - E_n (\tilde{p}_z) $ on $ \tilde{p}_z $ can be
neglected. Consequently, a real part of the conductivity tensor
for this group of electrons can be written in the form:
\begin{equation}
Re \, \sigma_{ii} (\omega) \approx N \sum_{n,m}
\frac{\sigma^{ii}_{nm}}{\frac{\tau^2}{\hbar^2}( E_n
(\tilde{p}_{z0}) - E_m (\tilde{p}_{z0}) - \hbar\omega)^2 + 1 } ,
\label{sigma-series}
\end{equation}
where $ \, N \, $ is the number of self-intersection points of the
Fermi surface, $ N = 2 $ for the model (\ref{spektr-abrikosov}).
The magnitudes $ E_{n,m} $ are evaluated by the expression
(\ref{En-full}) given at the self-intersection point of the Fermi
surface $ \tilde{p}_{z0} = p_{z0} \cos \theta $, $ p_{z0} =
\frac{\hbar}{a_z}  \arccos \left( -\frac{\epsilon_F}{2t} \right)$.
The approximate value of the conductivity tensor differs from the
exact value, which takes into account all groups of electrons, by
a correction amendment $ \Delta \sigma << \frac{t}{\hbar^2
\omega^2 \tau} \frac{e^2}{a_z} $, which is negligible in
comparison with the characteristic values of the conductivity
tensor (\ref{sigma-series}) due to the left side of the inequality
(\ref{usl-T}).

Each contribution $ \sigma^{ii}_{nm} $ is determined only by the
transitions between the Landau levels with numbers n,m and
correspond to the maximum of the conductivity  $ Re \; \sigma_{ii}
(\omega) $ at the resonance frequency $ \hbar\omega = \epsilon_n -
\epsilon_m $ if the mutual overlap of the resonance peaks is
omitted. In the region delimited by (\ref{omega-usl}) only
resonance frequencies corresponding to electron-hole transitions
of the charge carriers are found where their energy spectrum are
approximately linear. For the harmonics of the quantum cyclotron
resonance with not too high order numbers
\begin{equation}
|n|,|m| << 1/ (k^2 \eta^2), \qquad k = |n| - |m|, \label{usl-lin}
\end{equation}
that means the linear approximation for the electron energy
spectrum in the calculation of the matrix elements of the velocity
operator, the contributions $ \sigma^{ii}_{nm} $ can be written in
the form:
\begin{equation}
\sigma^{ii}_{nm} = \frac{2e^3 B \tau \cos \theta}{(2\pi\hbar)^2 c
| v_\bot |} | v^i_{nm} |^2 , \label{sigma-general}
\end{equation}
here $ v_\bot, $ (\ref{const-theta-2}) and $ v^{i}_{nm} $
(\ref{vpm})  are determined by the linear energy spectrum
(\ref{spektr}) with the parameters (\ref{const-theta-2}, \ref{En})
evaluated at $ {\bf p } = (0,0,p_{z0}) $.
The phase independence of the quantum oscillations of the
conductivity tensor (\ref{sigma-series},\ref{sigma-general}) on
the Fermi energy under the conditions of quantum cyclotron
resonance  and the absence of temperature damping of the
oscillations at not so high temperatures, when the electron-phonon
scattering can be neglected, are associated with the fact that in
case of the linear energy spectrum (\ref{spektr}), the cyclotron
frequency of the charge carriers of Dirac type depends strongly on
the number of Landau levels, but is practically the same for
charge carriers which have different momentum component along the
magnetic field direction (see the part 4.C of the Ref.\cite{LAS}).
In case of a tilted magnetic field we can can neglect the
difference between the linear relation (\ref{spektr}) and the
exact energy spectrum within the limits of the temperature
smearing of the Fermi level when inequality (\ref{usl-T}) is
satisfied. In a quantized magnetic field orthogonal to the layers,
the possibility of observing a high-temperature effect for the
conductors of the graphite family was predicted in
Ref.\cite{KozlovMedina}, where preliminary evaluations were
provided.
\begin{figure}
\begin{center}
\includegraphics[width=\linewidth]{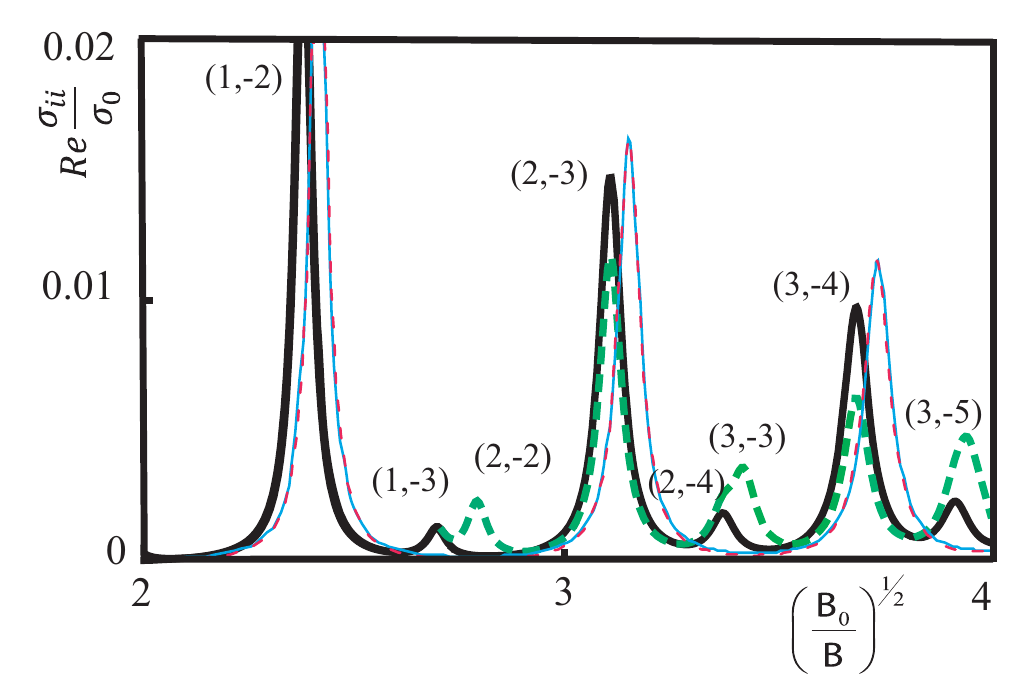}
\caption{\label{fig:ris1n} (Color online) The dependence of the
conductivity tensor $ Re \sigma_{xx} $ (continuous line) and $ Re
\sigma_{\tilde{y}\tilde{y}} $ (dashed line) on the magnetic field
B for a constant electromagnetic field frequency, when $ \tan
\theta = 0 $ (thin line) and $ \tan \theta = 2/3 $ (thick line).
The constants used on the axes labels are $ \sigma_0 = 2 e^2 /
(\hbar a_z)$, $ B_0 = \frac{c \hbar \omega^2}{2 e v_{\|}^2}$. The
parameter values used are $ t/(\hbar \omega) = 8 $, $ \epsilon_F
/(\hbar \omega) = 1 $, $ v_{\bot}/v_{\|} = 0.3 $, $ \omega\tau =
100 $. The pairs of numbers $ (n,m) $ determine the numbers of a
pair of Landau levels forming the given resonance peak.}
\end{center}
\end{figure}

\begin{figure}
\begin{center}
\includegraphics[width=\linewidth]{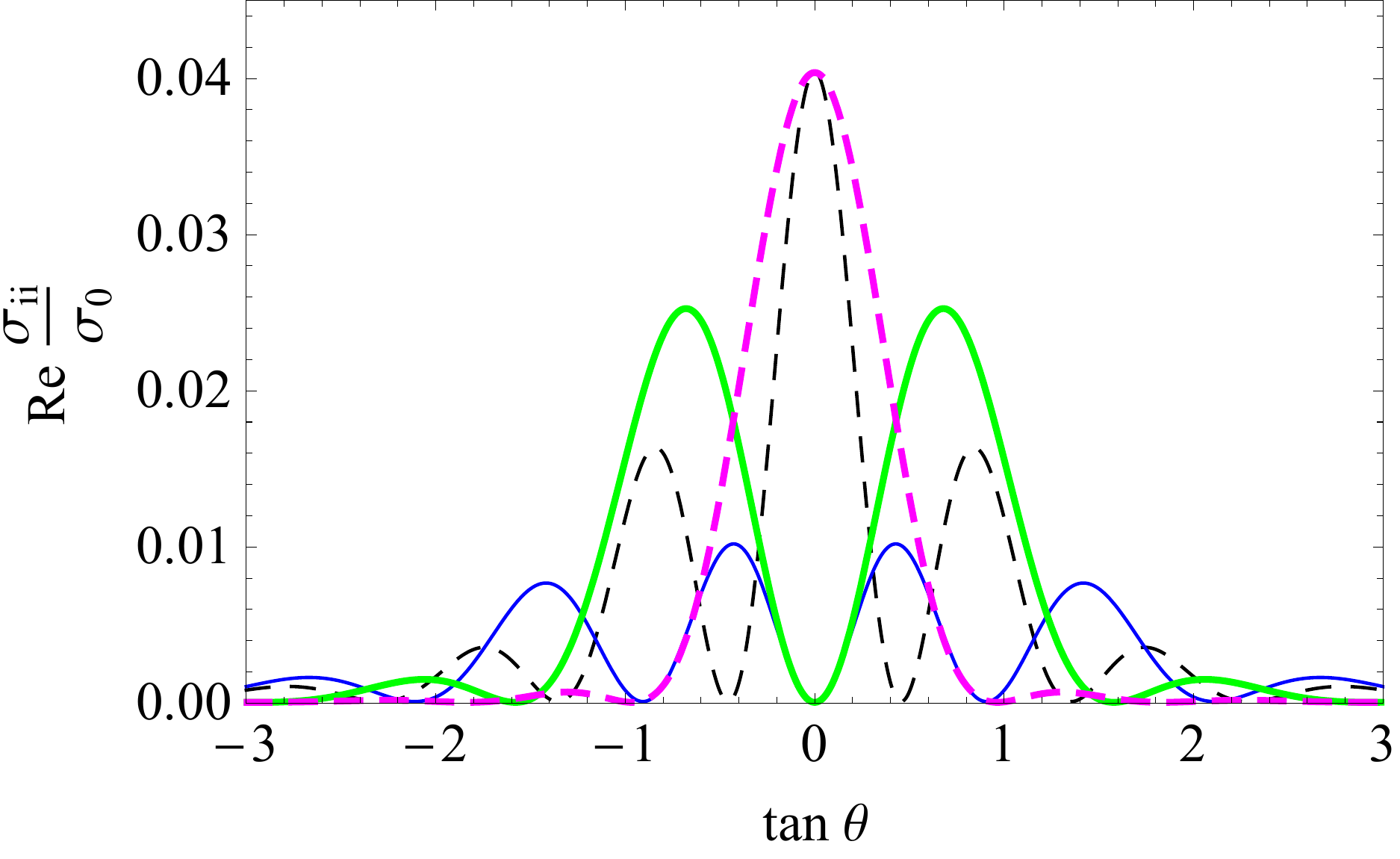}
\caption{\label{fig:ris2n} (Color online) The angular dependence
of the maximum value of $ Re \sigma_{\tilde{y} \tilde{y}} $ (thick
line) and $ Re \sigma_{xx} $ (thin line) near the resonance $
(3,-4) $ (dashed line) and $ (2,-4) $ (continuous line),
normalized by the constant $ \sigma_0 = 2 e^2 / (\hbar a_z)$, for
a fixed value of the electromagnetic wave frequency $ \omega $.
The parameter values used are $ t/(\hbar \omega) = 7 $, $
\epsilon_F /(\hbar \omega) = 1.8 $, $ v_{\bot}/v_{\|} = 0.3 $, $
\omega\tau = 300 $.}
\end{center}
\end{figure}

Fig.\ref{fig:ris1n}  shows the behavior of the diagonal components
of the conductivity tensor as a function of the magnetic field
magnitude for the fixed magnetic field tilt $ \theta $ and
frequency $ \omega $. The relation is numerically built taking
into account all the groups of electrons, $ -\frac{\pi\hbar}{a_z}
< p_z \leq \frac{\pi\hbar}{a_z} $ using the expressions
(\ref{spektr-abrikosov},\ref{const-theta-2},\ref{kin-ur},\ref{En},\ref{vpm}).
The similar dependence built using the approximate expression
(\ref{sigma-series}) with the same values of the parameters is not
visually different from that one shown in the figure.
The pair of numbers $ (n,m) $ at each peak in Fig.\ref{fig:ris1n}
and its reflection $ (-m,-n) $ correspond to the Landau level
numbers and the most important contributions to $ \sigma^{ii}_{nm}
$ (\ref{sigma-general}) forming the shown resonant peak. In a
tilted magnetic field besides the main peaks $ |n| - |m| = \pm 1
$, which determine the representation of the cyclotron resonance
for $ \theta = 0 $, higher harmonics are added. In these
harmonics, the amplitude of the sufficiently high peaks of the
pair $ (n,m) $ shows oscillations as a function of the angle $
\theta $. Fig.\ref{fig:ris2n} shows the angular dependence of a
resonance peak amplitude for a fixed frequency $ \omega $.

The physical mechanism of these oscillations can be explained as
follows. The energy of the conduction electron $ \epsilon
(\tilde{p}_x, \tilde{p}_y) $ in Larmor orbit's plane $ \tilde{p}_z
= const $ can be described using the anisotropic Dirac cone model
(\ref{spektr}). It is well known, that the corresponding
wavefunctions in a quantized magnetic field, which differ only by
their Landau level number n, can be expressed through the
Hermitian functions (\ref{psi}) with shifted center $ X_n  $
(\ref{Xn}) which magnitude depends only on the Landau level
number. Hence, when $ |n|, |m|
>> 1 $ the expressions for the component of the
velocity operator $ v^{\tilde{x},\tilde{y}}_{n,m} $  contain the
product of oscillating functions having a phase shift caused by
the difference $ X_n - X_m \neq 0 $, which depends on the
magnitude and direction of the magnetic field $ {\bf B} $. Their
interference leads to the oscillatory dependence $
v^{\tilde{x},\tilde{y}}_{n,m} $ (\ref{vxyshort}), and, therefore,
to  the oscillations of the conductivity tensor component.

\begin{figure*}
\begin{center}
\begin{minipage}[h]{0.49\linewidth}
(a) \\ \center{\includegraphics[width=\linewidth]{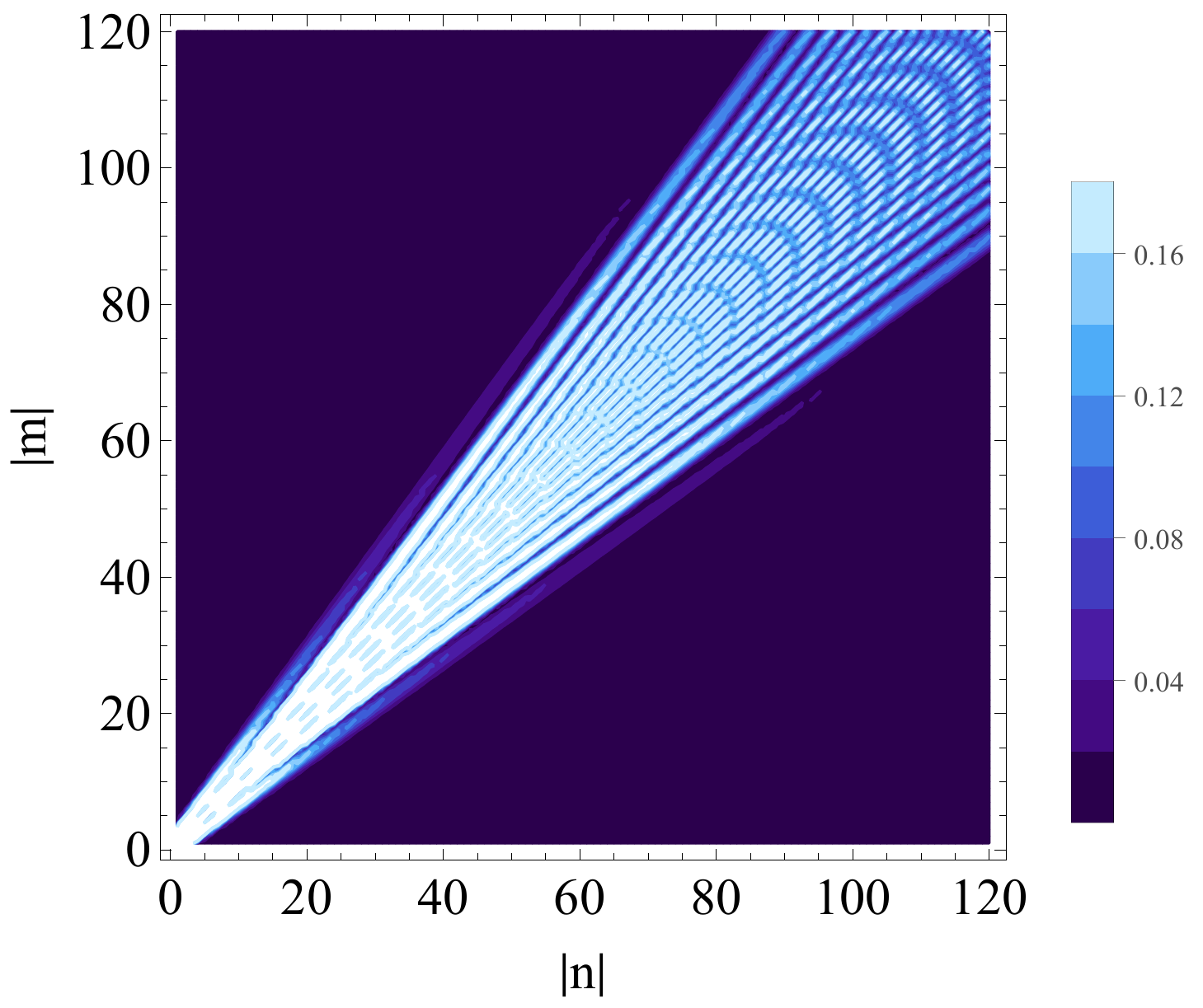}}
\end{minipage}
\hfill
\begin{minipage}[h]{0.49\linewidth}
(b) \\ \center{\includegraphics[width=\linewidth]{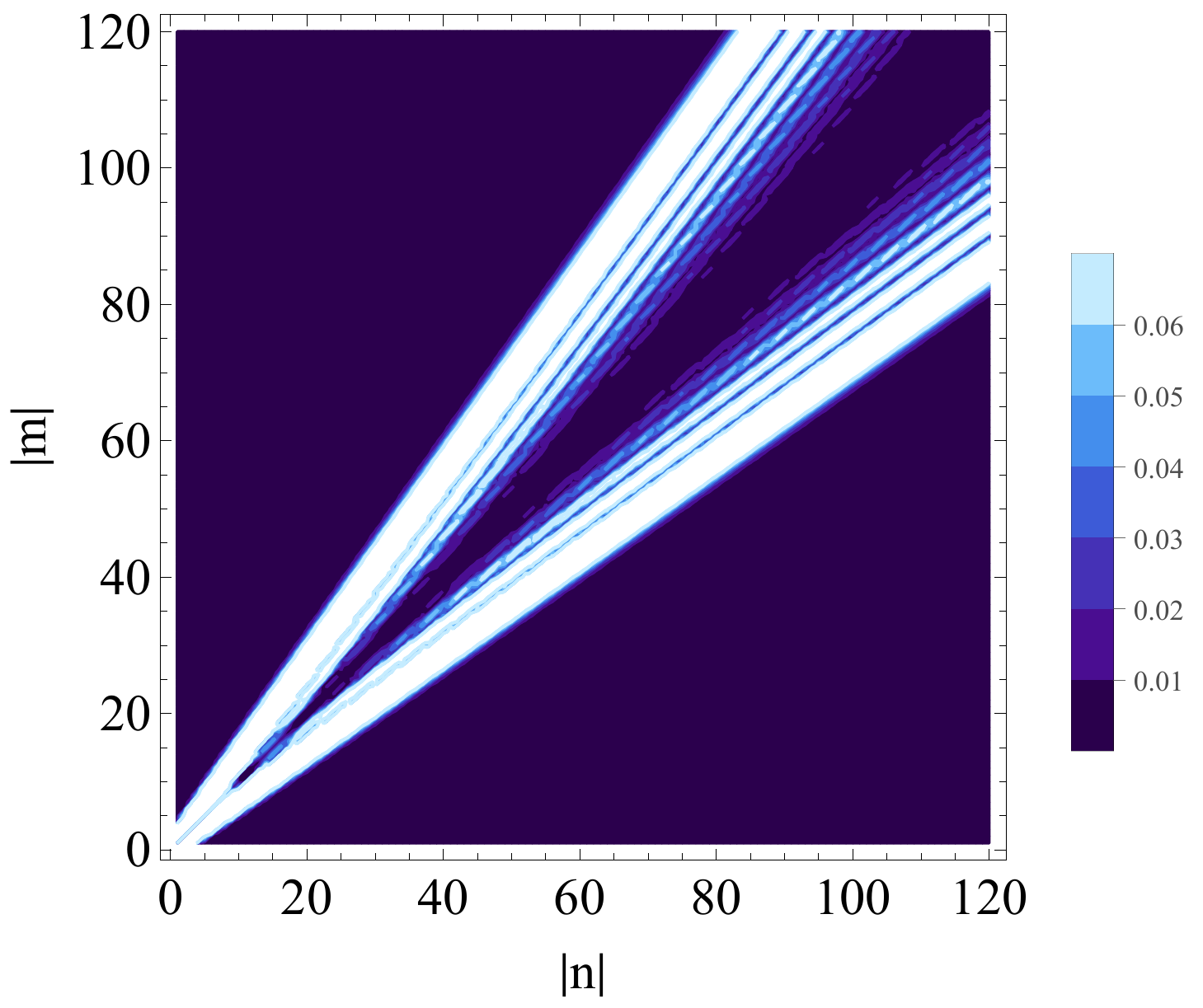}}
\end{minipage}
\caption{\label{fig:weer} (Color online) The dependence of the
contributions $ \sigma^{\tilde{y}\tilde{y}}_{nm} $ (a) and $
\sigma^{xx}_{nm} $ (b), normalized by $ \sigma_{\Sigma} $
(\ref{sum-rule}),  that determine the conductivity tensor
components (\ref{sigma-series})  under the resonance conditions $
\hbar\omega = \epsilon_n - \epsilon_m $ at the fixed magnitude and
direction of the magnetic field, $ \tan \theta = 0.3 $, $
v_{\bot}/v_{\|} = 0.3 $, on the numbers $ (n,m) $, which determine
the resonances caused by electron-hole transitions.}
\end{center}
\end{figure*}

The representation of the oscillations of the conductivity tensor
(\ref{sigma-general}) would be clearer if we use asymptotic
expressions for the velocity operator.  One can admit  that for
the velocity operator components $ v^{\tilde{x},\tilde{y}}_{n,m} $
\cite{Sari2}, which are related to the electron-hole transitions $
sign(n) \neq sign(m) $ and limited by the condition
\begin{equation}
|n| - |m| << 1/\eta, \sqrt{|n|} , \label{vxy-usl}
\end{equation}
the known asymptotic expression $ L^\alpha_j (x) \approx J_\alpha
(2\sqrt{jx}) $ can be applied yielding the components' simple
asymptotic expression:
\begin{equation}
v_{nm}^{\tilde{y}} \approx v_{0} \lambda^2  J^{\, '}_{|k|} (4 \eta
l) , \; \; v_{nm}^{\tilde{x}} \approx i v_{0} \lambda \alpha
\frac{k}{4 \eta l}
 J_{|k|} (4 \eta l) ,
\label{vxyshort}
\end{equation}
where $ k = |m| - |n|, \qquad l = min(|m|,|n|) $ and $ J^{\, '}_k
(x) $ is the derivative of the Bessel function.
The asymptotic form of the velocity operator components is
insignificantly different from (\ref{vpm}) for the physical
picture of the oscillations phase shift and does not account for
the overwhelming multiplier $ \exp(-2 \eta^2 l) \approx 1 $ when $
\eta << \frac{1}{\sqrt{|n|}} $, as the condition (\ref{omega-usl})
holds true.
A more accurate, though awkward, the asymptotic expansion for
associated Laguerre polynomials $ L_n^a (z) $, in particular, for
the oscillatory behavior of the region $ 0 < z < 4 n + 2 (a + 1)
$, can be found in Ref.\cite{Temme}.
The expressions (\ref{vxyshort}) maintain the physical structure
of the  velocity operator oscillations. This is the way the
asymptotic value of  $ v_{nm}^{\tilde{x},\tilde{y}} $
(\ref{vxyshort}),  as well as its exact expression (\ref{vpm}),
will be significantly different from zero only in the region of $
|\eta| < \frac{|k|}{|n| + |m|} $, and exponentially little beyond
it (this condition is easier to obtain using the WKB approximation
in conjunction with the method of a stationary phase).
Fig.\ref{fig:weer}  shows the dependence of the resonance peaks
amplitude(\ref{sigma-general}) on  m and n numbers. The
oscillations of values $ \sigma_{ii}^{nm} $ are the result of the
anisotropy of the energy spectrum of Dirac type that is indirectly
confirmed by qualitative similarity of the given figure and Fig.2
(a, c) of the Ref.\cite{Sari}. While constructing
Fig.\ref{fig:weer} the exact expressions for the matrix elements
of the velocity operator $ v^i_{nm} $ for the energy spectrum
(\ref{spektr}) were used, however replacing them with the
approximate values (\ref{vxyshort}) describes correctly the
oscillation dependence of the peak amplitude in terms of
inequality (\ref{vxy-usl}).

\section{Sum rule}
The tilt angle  of the magnetic field $ \theta $, corresponding to
the condition $ \eta (\theta) \sim 1/l $, separates the two cases
of the quantum cyclotron resonance. At the smaller angles $ \theta
$, the quantum cyclotron resonance will be determined only by the
fundamental harmonics of $ k = |n| - |m| = \pm 1 $. For the larger
angles $ \theta $, in the frequency range (\ref{omega-usl}) a lot
of higher resonance harmonics will appear, while the amplitude of
the fundamental harmonics caused by electron-hole transitions $ n
+ m = \pm 1 $ falls sharply.
%
%{\color{red} The influence of transitions between electron and
%hole bands on the kinetic coefficients of graphene and related
%conductors is a natural analogue of the "Zitterbewegung" effect
%\cite{Katsnelson}. }

It may be noticed, that there is a kind of rule of conservation of
the resonance peaks of total amplitude, explaining the decrease of
the amplitude of the fundamental resonances during the appearance
of higher harmonics. Namely, for an arbitrary $\eta$ and $ n =
const $ the relation:
\begin{equation}
\sum_{m=-\infty}^\infty |v^{\tilde{x}}_{nm}|^2 = v_0^2 \alpha^2,
\qquad \sum_{m=-\infty}^\infty |v^{\tilde{y}}_{nm}|^2 = v_0^2 (1 +
\eta^2) , \label{v-sum}
\end{equation}
is valid, which follows directly from the properties of the Pauli
matrices  $ \sigma^2_{x,y} = 1 $. The expressions (\ref{v-sum})
remain valid when using the asymptotics (\ref{vxyshort}) and pass
to the known sum rule $ J^2_0 (x) + 2\sum_{n=1}^\infty J^2_n (x) =
1 $. From  the expressions (\ref{sigma-general},\ref{v-sum}) it
follows that the maximums of the cyclotron resonance peaks due to
the charge carriers near the Dirac singularity
(\ref{omega-usl},\ref{usl-lin}) obtained for the same values of
magnitude and direction of the magnetic field {\bf B} (with
different resonance frequencies $\omega$) satisfy the relationship
\begin{eqnarray}
\sum_{m=-\infty}^\infty  \sigma^{\tilde{x}\tilde{x}}_{nm} =
\alpha^2 \sigma_\Sigma , \qquad
\nonumber  \\
\sum_{m=-\infty}^\infty \sigma^{\tilde{y}\tilde{y}}_{nm} = (1 +
\eta^2) \sigma_\Sigma ,
\nonumber  \\
\sigma_\Sigma = \frac{2e^3 B \tau v_0^2 \cos \theta}{(2\pi\hbar)^2
c | v_\bot |} , \qquad \label{sum-rule}
\end{eqnarray}
in which summed contributions visually correspond to one of the
horizontal in Fig.\ref{fig:weer}.  So we have the following "magic
square rule" for a table built from the resonance peak amplitude
values of the conductivity $ \sigma_{nm} $ (\ref{sum-rule}): the
sums of all the elements in the rows $ (n=const) $ and columns $
(m=const) $ do not depend on their numbers and they are equal.

\section{Conclusions}
The found oscillatory dependence of the conductivity tensor
(\ref{sigma-series},\ref{sigma-general},\ref{vxyshort}) has a
quantum interference nature and is a consequence of the anisotropy
of the electron energy spectrum in Larmor orbit's plane, which
arises in a tilted magnetic field.
The amplitude of the resonance peaks satisfies the simple sum rule
or the "magic square rule", which follows directly from properties
of Pauli matrices.
The character of the oscillatory dependence is similar to those
observed in Ref.\cite{Sari} oscillations of the absorption
coefficient of the electromagnetic field for a two-dimensional
conductor of Dirac type with a natural anisotropy of the electron
energy spectrum, or in crossed electric and magnetic fields, as a
function of an electric field or the degree of deformation of the
conductor.
Unlike the two-dimensional case, in graphite family conductors the
degree of the Dirac cone anisotropy $ \eta $ can be modified by
simply changing the inclination angle of a quantized magnetic
field, which substantially facilitates the conditions for the
experimental observation of oscillatory phenomena that are related
to the Dirac cone anisotropy.

Providing that the charge carrier velocity in the plane of the
layers $ v_0 $ (\ref{spektr-abrikosov}) is close to its value in
graphene (see, for example, Tab.2 in Ref.\cite{sarma2011}), the
resonance frequency corresponding to the transition between zeroth
and the first Landau levels $ \omega \sim 5 \times 10^{13} \times
\sqrt{B [T]} $ [Hz], when the magnetic field is directed by the
normal to the layers, and decreases if magnetic field tilt angle $
\theta $ is increasing  according to the expression (\ref{En}).
Thus, the region of the resonance frequencies in
Fig.\ref{fig:ris1n} will be limited to the submillimeter and
infrared diapason. Although the model (\ref{spektr-abrikosov}) is
suitable for multilayers of graphene with a AA type of stacking of
the crystal lattice, but it can also be used for the description
of the physical properties of other anisotropic conductors with a
Dirac singularity in the electron energy spectrum,
 the characteristics of which may differ much from
the similar values in graphene, including the region of resonance
frequencies.
However, the observed oscillatory effect is not restricted by the
given model and can take place in different conductors with nodal
line in the energy spectrum.
In particular, the energy spectrum of graphite with AB type of
stacking is also characterized by a non-zero Berry phase
\cite{Mikitik} and has a local structure (\ref{spektr}) in the
proximity of self-intersection points of the Fermi surface
\cite{MacClure57,SW}. The purity of highly oriented pyrolytic
graphite (HOPG) gives the possibility of experimental observation
of higher harmonics of the quantum cyclotron resonance
\cite{Orlita} and the angular oscillations of the kinetic
coefficients in the frequency domain of the electromagnetic wave
of millimeter and infrared range (see, for example
Ref.\cite{goncharuk2012} and references therein). The
investigation of the angular oscillations of high-frequency
kinetic coefficients which are caused by the charge carriers of
Dirac type in graphite of AB type stacking is beyond the scope of
this article and will be presented in a separate paper.

The absence of an inversion center ($ \eta \neq 0 $) of the model
(\ref{spektr}) in Larmor orbit's plane, being the cause of the
oscillatory dependence on the matrix elements of the velocity
operator (\ref{vxyshort}), though does not lead to quantitative
changes of the  quantized energy spectrum (\ref{En}).
Therefore, the interference mechanism observed here may take place
in the kinetic coefficients, which are related to electron
transport phenomena (electrical conductivity, impedance) and at
the same time can not cause the magneto-angular oscillations of
the density of states and the related thermodynamic
characteristics of a conductor.
Naturally the effects specific to a Dirac anisotropic spectrum are
not limited to high-frequency transport phenomena. Thus the phase
transition of $ 3\frac{1}{2} $ kind in conductors with nodal lines
in the energy spectrum of charge carriers, which is sensitive to
the anisotropy of the Dirac electron energy spectrum, is described
in Ref.\cite{Mikitik2014}.

The magneto-angular oscillations in bilayer graphene predicted in
Ref.\cite{pershoguba} have a similar physical nature as they are
explained by interference of wave functions with the displacement
of the centers of Larmor orbits in the graphene neighboring
layers.
However, the effect leads to the occurrence of the magneto-angular
oscillations in the density of states of the electronic subsystem,
which differs it from the mechanism of oscillations appearance
Fig.\ref{fig:ris2n}. Also in contrast to oscillations of the
conductivity tensor
(\ref{sigma-series},\ref{sigma-general},\ref{vxyshort}), which
period of oscillations is determined by the ratio of the Fermi
velocities in directions perpendicular and parallel to the layers,
the overlap integral between the layers does not affect the phase
of the oscillations \cite{pershoguba}, although determines their
amplitude.
While working over the present article, we came across
Ref.\cite{tchumakov-PRL}, where type-II Weyl semimetals in a
tilted magnetic field were investigated and Landau quantization
was proved to be possible even in the given conductors for
magnetic field directions with the effective tilt $ \eta < 1 $.
The existence of a new type of angular oscillations of kinetic
coefficients for the conductors of the graphite family considered
in the presented work was announced in the abstract
\cite{FTT-2015}.

The authors express the gratitude to FINCYT and CONCYTEC of Peru
for financial support of this work.

\section*{Authors contributions}
The contributions of the three authors are equal.

\appendix
\numberwithin{equation}{section}
\section{Matrix elements of the velocity
operator}
The eigenvalues $ \epsilon_n $ and wave functions $ \varphi_\nu (
{\bf r} ) $ of the Hamiltonian (\ref{spektr}) in a quantized
magnetic field with the gauge $ A = (0,By,0) $ have the form:
\begin{equation}
\epsilon_n = v_0 \; sign(n) \sqrt{2 \frac{eB\hbar}{c} \lambda^3 \,
\alpha \, |n|} ,  \label{En}
\end{equation}
\begin{eqnarray}
\varphi_\nu ( x , y ) = \frac{(\alpha\lambda)^{1/4}}{2(2\pi \hbar)
\sqrt{a_H}} \sqrt{\frac{1 + \delta_{0,n}}{1+\lambda}} \qquad
\qquad \nonumber  \\   \times
\exp(\frac{i}{\hbar} P_y y)  \left\{ \left[ \begin{array}{c}
{i\eta} \\ {1+\lambda} \end{array} \right] h_{|n|}
(\frac{\sqrt{\alpha\lambda}}{a_H}(x + X_n)) \right. \qquad
\nonumber  \\
\left. - \left[ \begin{array}{c} {i (1 + \lambda)} \\
{\eta} \end{array} \right] sign \, (n) \, h_{|n| - 1}
(\frac{\sqrt{\alpha\lambda}}{a_H}(x + X_n)) \right\} , \qquad
\label{psi}
\end{eqnarray}

where $ \nu = (n, P_y) $ is the complete quantum index set, {\bf
P} is the canonical momentum,  the magnetic length
\begin{equation}
a_H = \sqrt{\frac{c\hbar}{eB}}, \qquad \lambda = \sqrt{1 - \eta^2}
, \label{lambda}
\end{equation}
the negative values of the Landau level numbers correspond to
holes in the energy spectrum of charge carriers,
\begin{equation}
X_n = a_H \,  \eta \, sign \, (n) \,
\sqrt{\frac{2|n|}{\alpha\lambda}} - \frac{cP_y}{eB} \label{Xn}
\end{equation}
is the centre of Larmor orbit of the conduction electrons,
\begin{equation}
h_n (\xi) = \frac{1}{\sqrt{2^n \sqrt{\pi} n!}} \exp (-\xi^2 / 2)
H_n (\xi)
\end{equation}
is the solution of the dimensionless harmonic oscillator problem,
$  H_n (\xi) $ is the n-th Hermite polynomial and $ \delta_{0,n} $
is the Kronecker symbol.
It is considered that the contribution containing sign(n),
(\ref{psi}) is equal to zero  when $ n = 0 $.

 The matrix elements of the velocity operator components have the
form
\begin{equation}
v^{y}_{nm} = \lambda (\Phi_{nm} + \Phi_{mn}), \qquad v^{x}_{nm} =
i \alpha (\Phi_{nm} - \Phi_{mn})  \label{vpm}
\end{equation}
where
\begin{eqnarray}
\Phi_{nm} = v_0 \lambda \sqrt{\frac{|n|}{2^{|m|-|n|+1}}}
\sqrt{\frac{|m|!}{|n|!}} \Delta_{nm}^{|n|-|m|-1}
e^{-\Delta_{nm}^2}
\nonumber \\ \times
L_{|m|}^{|n|-|m|-1} (2\Delta_{nm}^2) sign(n) \qquad
\end{eqnarray}
\begin{equation}
\Delta_{nm} = \frac{\eta}{\sqrt{2}} (sign(n)\sqrt{|n|} -
sign(m)\sqrt{|m|}) , \label{Deltanm}
\end{equation}
when $ n \neq 0 $ and $ \Phi_{0m} = 0 $.

The expressions similar to (\ref{En},\ref{psi},\ref{vpm}) are
given in a series of works (for example, see
\cite{Sari,Morinari}). In particular, the expression (\ref{vpm})
corresponds to the formulae (A1-A2) of Ref.\cite{Sari2} where the
value of the parameter $ \alpha = 1 $, if the dependence on the
latter is considered by simple coordinate transformation $ y' = y
, \ x' = x / \alpha $.

%\newpage

\end{document}